\documentclass[preprint,prl,oneside,superscriptaddress,showpacs,floatfix,endfloats*]{revtex4}

\usepackage{amsmath}
\usepackage{amssymb}
\usepackage{graphicx}
\usepackage{dcolumn}
\usepackage{bm}
\usepackage{color}

\def\m#1{\mathrm{#1}}
\def\Eq#1{(\ref{eq:#1})}
\def\d{\mathrm{d}}

\def\Fig#1{\ref{fig:#1}}

\def\epsilon{\varepsilon}
\def\theta{\vartheta}
\def\rho{\varrho}
\def\Int#1#2#3{\int\limits_{#1}^{#2}\!\mathrm{d}{#3}\;}

\begin{document}


\title{Electrostatic interactions in critical solvents}

\author{Markus Bier}
\email{bier@is.mpg.de}
\affiliation
{
   Max-Planck-Institut f\"ur Intelligente Systeme, 
   Heisenbergstr.\ 3,
   D-70569 Stuttgart,
   Germany, 
   and
   Institut f\"ur Theoretische und Angewandte Physik,
   Universit\"at Stuttgart,
   Pfaffenwaldring 57,
   70569 Stuttgart,
   Germany
}

\author{Andrea Gambassi}
\affiliation
{
   SISSA --- International School for Advanced Studies and INFN,
   34136 Trieste,
   Italy,
   and 
   Laboratoire de Physique Th\'eoretique et Hautes Energies,
   UMR 7589,
   Universit\'e Pierre et Marie Curie ---
   Paris VI, 75252 Paris Cedex 05,
   France
}

\author{Martin Oettel}
\affiliation
{
   Institut f\"ur Physik, 
   Johannes-Gutenberg-Universit\"at Mainz,
   WA 331,
   55099 Mainz, 
   Germany
}

\author{S.\ Dietrich}
\affiliation
{
   Max-Planck-Institut f\"ur Intelligente Systeme, 
   Heisenbergstr.\ 3,
   D-70569 Stuttgart,
   Germany, 
   and
   Institut f\"ur Theoretische und Angewandte Physik,
   Universit\"at Stuttgart,
   Pfaffenwaldring 57,
   70569 Stuttgart,
   Germany
}

\date{April 29, 2011}

\begin{abstract}
The subtle interplay between critical phenomena and electrostatics is investigated by considering
the effective force acting on two parallel walls confining a near-critical binary liquid mixture
with added salt.
The ion-solvent coupling can turn a non-critical repulsive electrostatic force into an attractive
one upon approaching the critical point.
However, the effective force is eventually dominated by the critical Casimir effect, the universal
properties of which are not altered by the presence of salt.
This observation allows a consistent interpretation of recent experimental data.
\end{abstract}


\maketitle


\paragraph{Introduction.} Effective interactions among surfaces in contact with fluid media play
a central role for a variety of topical fields in soft and condensed matter physics, cell biology,
colloid and surface science, and nanotechnology.
Since many relevant fluid media contain polar liquids such as water, their confining surfaces
acquire an electric charge due to ion association or dissociation.
As a result, electrostatic forces are expected to contribute significantly to the interactions.
In addition, a strong and highly temperature-sensitive solvent-mediated effective force arises
upon approaching critical points.
This critical Casimir force has recently been reported for a single colloidal particle close to
a wall and immersed in a binary liquid mixture near its critical demixing point 
\cite{Hertlein2008}.
This force is expected to play also a role for the aggregation of colloidal suspensions
\cite{Bonn2009,Gambassi2010}.
Motivated by recent experiments, here we investigate the interplay between electrostatic and
critical Casimir forces, which turns out to be responsible for rather unexpected effects in 
binary liquid mixtures with added salt.
Colloids dispersed in such a solvent have been reported \cite{Bonn2009} to aggregate at a
temperature difference from the critical demixing point which increases upon increasing the
ionic strength, i.e., the screening of the electrostatic forces.
This observation has been confirmed experimentally also for a single colloid near a wall
\cite{Nellen2011}.
While it was originally argued \cite{Bonn2009} that the aggregation could be completely
explained in terms of a simple superposition of the critical Casimir and screened Coulomb forces
(see, however, Ref.~\cite{Gambassi2010}), subsequent experimental results challenge this picture:
An \emph{attractive} colloid-wall interaction has been observed within a suitable temperature
range even though \emph{both} the electrostatic and the critical Casimir force are expected
to be separately \emph{repulsive} under these experimental conditions \cite{Nellen2011}.
This points towards an important and yet unexplored aspect of the coupling between 
electrostatics and the critical fluctuations of the medium.
Certain features of ion-solvent coupling near critical points were investigated in the past,
such as the possibility of a micro-heterogeneous phase \cite{Nabutovskii1980} and the
influence of criticality onto the solubility of ions \cite{Onuki2004}.
However, the complementary point of view, i.e., the influence of ions onto the critical
fluctuations of a solvent and therefore onto the critical Casimir effect has not yet been
studied.
We present a minimalistic but sufficiently enriched theory which explains the aforementioned
unexpected experimental results.
This contribution is intended to initiate a cross fertilization between research areas which so
far focused separately on the critical Casimir effect in salt-free systems or on
electrostatic interactions in fluctuation-free solvents.
We expect that understanding --- and thus being able to use on purpose --- the coupling 
between the critical Casimir effect and electrostatics provides a key to push forward
the topical fields mentioned above.
Our findings actually reach beyond the realm of soft matter because the 
counterintuitive behavior en route to universality discussed here is expected 
to be paradigmatic for many branches of physics in which generically critical and 
non-critical fields are coupled.


\paragraph{Model.} In a three-dimensional ($d=3$) Cartesian coordinate system, we consider
two parallel walls located at positions $\widetilde{z}=0$ and $\widetilde{z}=\widetilde{L}>0$,
respectively, with the space in between being filled  by a binary liquid mixture.
In this solvent cations ($+$) and anions ($-$) are dissolved.
The solvent particles are all assumed to be of equal size with volume $\widetilde{a}^3>0$ whereas the
ions are considered point-like.
At the dimensionless position $z:=\widetilde{z}/\widetilde{a}$, the number densities of the 
solvent components are given by $\phi(z)\widetilde{a}^{-3}$ and $(1-\phi(z))\widetilde{a}^{-3}$
with $0\leq\phi\leq1$, whereas the densities of the cations and anions are given by 
$\rho_+(z)\widetilde{a}^{-3}$ and $\rho_-(z)\widetilde{a}^{-3}$, respectively.
The walls carry surface charge densities $\sigma_0e\widetilde{a}^{-2}$ at $z=0$ and 
$\sigma_Le\widetilde{a}^{-2}$ at $z=L:=\widetilde{L}/\widetilde{a}$, where $e$ is the elementary
charge.
The composition $\phi(z)$ couples to surface fields $h_0$ at $z=0$ and $h_L$ at $z=L$, where
$h_{0,L}>0$ ($<0$) leads to a preferential adsorption of the solvent component with $\phi=1$ 
($=0$).
The equilibrium profiles $\phi$, $\rho_+$, and $\rho_-$ minimize the approximate grand potential
density functional $k_BT\Omega[\phi,\rho_\pm]$,
\begin{eqnarray}
   && \frac{\Omega[\phi,\rho_\pm]}{A} = \Int{0}{L}{z} 
      \bigg\{\omega_\m{sol}(\phi(z)) + \frac{\chi(T)}{6}\phi'(z)^2 
      \nonumber\\
   && \hphantom{MM} 
      + \sum_{i=\pm}\Big[\omega^{(i)}_\m{ion}(\rho_i(z)) + \rho_i(z)V_i(\phi(z))\Big]
      \label{eq:df}\\
   && \hphantom{MM}  
      + 2\pi\ell_BD(z,[\rho_\pm])^2\bigg\} - h_0\phi(0) - h_L\phi(L),
      \nonumber
\end{eqnarray}
with $\omega_\m{sol}(\phi)=\phi(\ln\phi-\mu_\phi) + (1-\phi)\ln(1-\phi) + \chi(T)\phi(1-\phi)$
and $\omega^{(\pm)}_\m{ion}(\rho_\pm)=\rho_\pm(\ln\rho_\pm - 1 - \mu_\pm)$ as
the grand potential bulk densities of the solvent and the $\pm$-ions, respectively.
Here $k_BT$ is the thermal energy, $A\widetilde{a}^2$ is the area of one wall, $\mu_\phi k_BT$
and $\mu_\pm k_BT$ are the chemical potentials of the solvent composition and the $\pm$-ions, 
respectively, and $\ell_B\widetilde{a}=e^2/(4\pi\epsilon k_BT)$ is the Bjerrum length for a
uniform permittivity $\epsilon$; a $\phi$-dependent permittivity \cite{Samin2011} corresponds to
modified surface fields $h_{0,L}$ \cite{Bier2011}.
The (temperature-dependent) Flory-Huggins parameter $\chi(T)>0$ describes the solvent-solvent 
interaction, which leads to a phase separation in the range $\chi(T)\geq\chi(T_c)$; the gradient
term $\propto\phi'(z)^2$ accounts for the spatial variation of the solvent composition
\cite{Cahn1958}.
The ion-solvent interaction is given by the effective ion potential $k_BTV_\pm(\phi)$ due to
the solvent, with $V_\pm(\phi) = -\ln(1-\phi(1-\exp(-f_\pm)))$ and where
$f_\pm k_BT$ is the \emph{difference} of the bulk solvation free energies of a $\pm$-ion in 
solvents with $\phi=1$ and $\phi=0$.
This expression of $V_\pm(\phi)$ leads to a bulk phase diagram with a critical point 
which is shifted towards larger values of $\chi(T_c)$ as the salt concentration
increases \cite{Bier2011}.
$V_\pm(\phi)$ is an improvement of the standard approximation $\phi f_\pm$ (which it 
reduces to for $f_\pm \ll 1$) because for $f_\pm\gtrsim1$ the latter leads to multiple critical points
\cite{Bier2011} which are, however, not observed experimentally.
For $f_\pm\to\infty$ the ion potentials $V_\pm(\phi)$ reduce to $-\ln(1-\phi)$ which
describes the entropy loss and thus free energy increase due to the insolubility of ions in the
solvent component with $\phi=1$.
The electric displacement $D(z,[\rho_\pm])e\widetilde{a}^{-2}$ in Eq.~\Eq{df} fulfills Gauss'
law with fixed surface charges \cite{Russel1989}: $D'(z,[\rho_\pm])=\rho_+(z)-\rho_-(z), 
D(0,[\rho_\pm])=\sigma_0, D(L,[\rho_\pm])=-\sigma_L$.
Note that $D(z,[\rho_\pm])$ is generated by the $\pm$-ions and the surface charges
$\sigma_{0,L}$, independent of $\phi$.
Within the present model, ions interact with the walls only electrostatically.

In order to highlight the effect of the ion-solvent coupling we focus on an approximate 
grand potential functional for the solvent composition alone, which is obtained by expanding
$\Omega[\phi,\rho_\pm]$ in Eq.~\Eq{df} in terms of the order parameter $\varphi:=\phi-\phi_b$
and the ion density differences $\Delta\rho_\pm:=\rho_\pm-I$ retaining quadratic 
contributions as well as terms proportional to $\varphi^3$ and $\varphi^4$.
Here $\phi_b$ and $I=\rho_{\pm b}$ denote the bulk solvent composition and the bulk ionic
strength, respectively, corresponding to the chemical potentials $\mu_\phi$ and $\mu_\pm$.
The minimization with respect to $\Delta\rho_\pm$ leads to linear, analytically solvable 
Euler-Lagrange equations for $\Delta\rho_{\pm\m{eq}}(z,[\varphi])$, which are functionals of 
$\varphi$.
Inserting these solutions into Eq.~\Eq{df} one obtains a Ginzburg-Landau-type functional
\begin{eqnarray}
   && \frac{\mathcal{H}[\varphi]}{A} = \Int{0}{L}{z} 
      \Big\{U(z)\varphi(z) + \frac{t(T)}{2}\varphi(z)^2
      \nonumber\\
   && \hphantom{MM}+ \frac{g}{24}\varphi(z)^4 
                   + \frac{\chi(T)}{6}\varphi'(z)^2\Big\}
      \label{eq:effdf}\\
   && \hphantom{MM}- h_0\varphi(0) - h_L\varphi(L) + W(L) + \mathcal{O}((\Delta\gamma)^2)
      \nonumber
\end{eqnarray}
with the temperature-like variable $t(T):=1/\phi_b + 1/(1-\phi_b) - 2\chi(T)$.
Here we assume that the mixture is at its critical composition such that there is no 
$\varphi^3$-term.
The electrostatic effects are contained in the coupling 
$g := 2/\phi_b^3 + 2/(1-\phi_b)^3 + 6I(\gamma_+^4+\gamma_-^4), \gamma_\pm:=V'_\pm(\phi_b)$,
as well as in an ``external'' field generated by the surface charges $\sigma_{0,L}$:
\begin{eqnarray}
   && U(z) := -\frac{\kappa\Delta\gamma}{2(1-\exp(-2\kappa L))}
      \nonumber\\ 
   && \hphantom{MM}\times\big[(\sigma_0+\sigma_L\exp(-\kappa L))\exp(-\kappa z) +
      \label{eq:U}\\
   && \hphantom{MM\times\big[}(\sigma_L+\sigma_0\exp(-\kappa L))\exp(-\kappa(L-z))\big]
      \nonumber
\end{eqnarray}
with the Debye screening length $\kappa^{-1}=(8\pi\ell_BI)^{-1/2}$ and 
$\Delta\gamma:=\gamma_+-\gamma_-$.
The ``direct'', i.e., solely ion-mediated, electrostatic interaction between the walls
is given by $W(L) :=(4\pi\ell_B/\kappa)P(\kappa L, \sigma_0, \sigma_L)$ where
\begin{equation}
   P(x,y_0,y_L) := \frac{2y_0y_L + (y_0^2+y_L^2)\exp(-x)}{2\sinh(x)}.
   \label{eq:G}
\end{equation}
The ion-solvent coupling affects the critical point only at order $\mathcal{O}((\Delta\gamma)^2)$
\cite{Bier2011}.

Upon approaching the critical point the dimensionless bulk correlation length 
$\xi = \widetilde{\xi}/\widetilde{a} = \sqrt{\chi(T)/(3t(T))}$, which characterizes the 
exponential decay of the two-point correlation function, diverges.
Accordingly, on the scale $\xi$, $U(z)$ is localized at the boundaries and therefore it
merely modifies the surface fields $h_{0,L}$.
Consequently $\mathcal{H}$ turns into a standard  $\varphi^4$-theory, which
describes the critical behavior of the Ising universality class \cite{Pelissetto2002}.
Thus, within the present model, electrostatic interactions do not affect
the \emph{universal} critical behavior of the solvent.

The effective wall-wall interaction is defined by $\widetilde{\omega}(L):=
\omega(L)k_BT\widetilde{a}^{-2}$ with $\omega(L) := (\mathcal{H}(L)-\mathcal{H}(\infty))/A$, 
where $\mathcal{H}(L)/A$ is the minimum of Eq.~\Eq{effdf}.
In general, for the critical contribution one has $\omega(L)=\vartheta(L/\xi)/L^{d-1}$
with a universal scaling function $\vartheta(x)$, which depends only on the relative
\emph{signs} of $h_{0,L}$ \cite{Krech1997}, with $\theta(x\to0)=\m{const}$ and 
$\vartheta(x\to\infty)= Cx^{d-1}\exp(-x)$, where $C$ is a universal, boundary-condition-dependent
constant \cite{Krech1997}.

For a sufficiently small bulk correlation length, i.e., if $\xi\ln\xi \ll L$, the term 
$\propto\varphi^4$ in Eq.~\Eq{effdf} can be neglected relative to the term $\propto\varphi^2$.
The resulting quadratic functional can be readily minimized and leads to the approximate
effective wall-wall interaction
\begin{eqnarray}
   && \omega(L) = -\frac{3\xi}{\chi(T)}P(L/\xi,h_0,h_L) 
      + \frac{4\pi\ell_B}{\kappa}P(\kappa L, \sigma_0, \sigma_L)
   \nonumber\\
   && \hphantom{MM}-\Delta\gamma\frac{3\kappa\xi^2}{\chi(T)}
      (Q_1(\kappa L,\kappa\xi)(h_0\sigma_L + h_L\sigma_0) +
      \label{eq:omega}\\
   && \hphantom{MM-\Delta\gamma\frac{3\kappa\xi^2}{\chi(T)}(}
       Q_2(\kappa L,\kappa\xi)(h_0\sigma_0 + h_L\sigma_L)) + \mathcal{O}((\Delta\gamma)^2)
      \nonumber
\end{eqnarray}
with the function, which is analytical for $y>0$,
\begin{equation}
   Q_k(x,y) := \frac{1}{y^2-1}\Big(\frac{y\exp(-kx/y)}{1-\exp(-2x/y)} - 
   \frac{\exp(-kx)}{1-\exp(-2x)}\Big).
\end{equation}

In Eq.~\Eq{omega} the term $\propto P(L/\xi,h_0,h_L)$ corresponds to the contribution of the surface
fields to the wall-wall interaction in the absence of ion-solvent coupling ($\Delta\gamma=0$)
whereas the term $\propto P(\kappa L, \sigma_0,\sigma_L)$ is the direct electrostatic wall-wall 
interaction.
The term $\propto(h_0\sigma_L+h_L\sigma_0)$ is the interaction ($\propto\Delta\gamma$) of the 
order parameter profile close to one wall ($\propto h_{0,L}$) with the polarization of the diffuse 
ion layer due to the surface charge on the \emph{opposite} wall ($\propto\sigma_{L,0}$).
This coupling between the order parameter and the ion density profiles is the central result of
the present analysis and has important consequences (see below). The analogous term 
$\propto(h_0\sigma_0+h_L\sigma_L)$ is small. 


\begin{figure}
   \includegraphics[width=8.6cm]{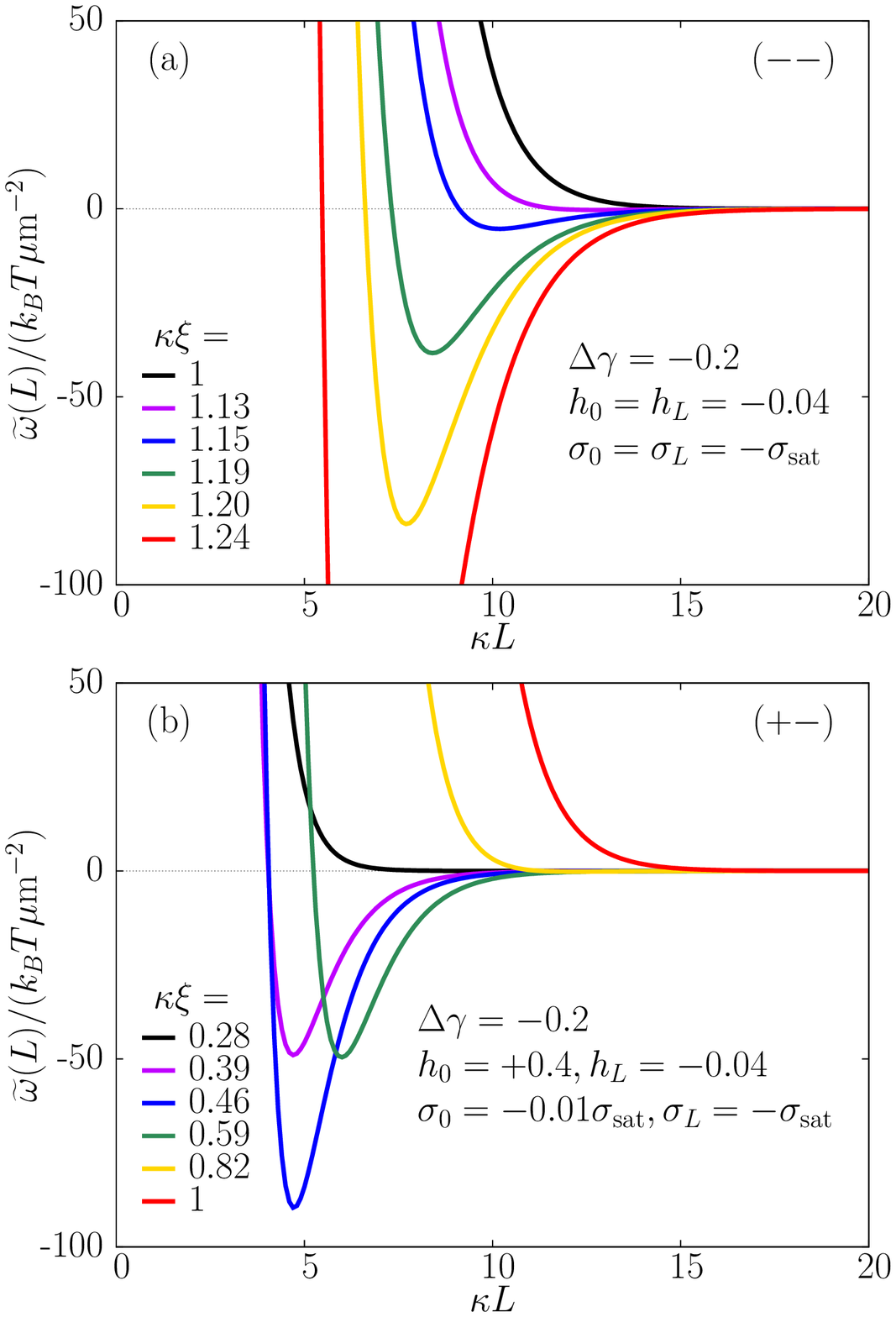}
   \caption{\label{fig:1}Effective wall-wall interaction potential $\widetilde{\omega}$ as a 
           function of the scaled wall separation $\kappa L$ and the scaled bulk correlation
           length $\kappa\xi$ for (a) $(--)$ and (b) $(+-)$ boundary conditions;
           $\sigma_\m{sat}=\kappa/(\pi\ell_B)$ is the saturation surface charge density.
           For symmetric $(--)$ boundary conditions $\widetilde{\omega}$ is repulsive far away
           from the critical point due to the direct electrostatic interaction between the 
           like-charged walls whereas the Casimir force gives rise to an increasing attraction 
           upon approaching the critical point.
           For antisymmetric $(+-)$ boundary conditions $\widetilde{\omega}$ is repulsive far
           away from as well as close to the critical point.
           Attraction ($\propto\Delta\gamma$) occurs in an intermediate
           temperature range due to the ion-solvent coupling induced by the difference
           between the solubility contrasts of cations and anions in the binary
           solvent.}
\end{figure}

\paragraph{Discussion.} In the following we discuss the experiments with colloids alluded to in
the Introduction. 
The predictions of the present model for two walls can be readily translated into those for the 
wall-sphere and sphere-sphere geometry by means of the Derjaguin approximation, which is 
applicable at separations much smaller than the sphere radii
\cite{Russel1989}.
It turns out that assuming additivity of Casimir and Coulomb forces, i.e., independence of
the order parameter from electrostatics, is in general insufficient to explain the experimental
observations, whereas the present model, which includes ion-solvent coupling, leads to a
consistent picture.

First we consider symmetric boundary conditions, $(h_0,h_L)=(-,-)$, for which the ion-solvent
coupling is masked by the strong direct electrostatic repulsion.
This situation has been investigated experimentally with a suspension of hydrophilic
spherical colloids in a water-oil mixture \cite{Bonn2009} as well as with a single 
hydrophilic colloidal sphere in a similar water-oil mixture near a hydrophilic glass wall 
\cite{Nellen2011}.
In the presence of salt aggregation \cite{Bonn2009} or strong wall-sphere attraction 
\cite{Nellen2011} has been observed upon approaching the critical point of the binary mixture already
several Kelvin away from the critical point.
Within the present model, this setting is described by $h_0=h_L<0$ and $\sigma_0=\sigma_L<0$ with
the composition $\phi$ expressed as the mole fraction of the non-aqueous component.
For a certain choice of parameters Fig.~\Fig{1}(a) displays the effective wall-wall interaction
potential $\widetilde{\omega}(L)$.
Since $\Delta\gamma \gtrless 0$ corresponds to $f_+ \gtrless f_-$, a negative ion-solvent coupling
strength $\Delta\gamma<0$ describes a salt the cations of which are slightly better soluble
in oil than the anions, which is expected because the oils used in the experiments, 
3-methylpyridine and 2,6-dimethylpyridine (2,6-lutidine), are Lewis bases \cite{Wade2006}.
The parameters used in Fig.~\Fig{1} correspond to a critical water-2,6-lutidine mixture
($\widetilde{a}=0.34\m{nm},\ell_B=2.82$) with $10\m{mM}$ salt ($1/\kappa=7.73$).
Far away from $T_c$ the effective wall-wall potential $\omega(L)$ exhibits a repulsion due to the
direct electrostatic wall-wall interaction.
Upon approaching $T_c$, $\omega(L)$ starts to develop an increasing attraction due to the 
critical Casimir effect. 
Since the change from repulsion to attraction occurs at $\kappa\xi\approx1$, the attraction sets
in only very close to the critical point if the ionic strength is small, whereas this change occurs
already considerably far away from the critical point if the ionic strength is large. 
Due to the strong direct electrostatic wall-wall interaction between hydrophilic walls, the
ion-solvent coupling does not qualitatively influence the effective wall-wall potential, so that
the assumption of additivity of critical Casimir and Coulomb forces \cite{Bonn2009,Gambassi2010}
is justified.

The situation is different for antisymmetric boundary conditions, $(h_0,h_L)=(+,-)$, as studied 
experimentally in Ref.~\cite{Nellen2011} using a single hydrophilic colloid near a
hydrophobic glass plate.
Repulsion is observed far away from as well as close to the critical point, whereas within an
intermediate temperature range a strong attraction is found.
The near-critical repulsion is readily understood in terms of the critical Casimir effect for
antisymmetric boundary conditions and the repulsion far away from the critical point is of 
electrostatic origin.
However, the attraction occurring in the intermediate temperature range cannot be explained
within a picture without ion-solvent coupling.
Figure~\Fig{1}(b) shows $\widetilde{\omega}(L)$ for a particular choice of non-symmetric 
surface fields $h_0>0, h_L<0$ and surface charge densities $\sigma_0,\sigma_L<0, 
|\sigma_0|\ll|\sigma_L|$ corresponding to a weakly charged hydrophobic wall.
Far from ($\kappa\xi\leq0.28$) and close to ($\kappa\xi\geq0.82$) the critical point 
$\widetilde{\omega}(L)$ is repulsive because in Eq.~\Eq{omega} the terms $\propto P$ dominate.
Upon increasing $\kappa\xi$ beyond $0.28$, i.e., en route towards $T_c$,
attraction occurs (see $\kappa\xi=0.39$), which for the chosen parameters is strongest around
$\kappa\xi=0.46$ and which weakens again closer to $T_c$ (see $\kappa\xi=0.59$).
This attraction is caused by the coupling ($\Delta\gamma<0$) between the order parameter profile
near the hydrophobic wall ($\propto h_0>0$) and the electrostatic potential due to the opposite, 
hydrophilic wall ($\propto \sigma_L<0$).
For this effect to take place it is essential that the hydrophobic wall is sufficiently weakly charged 
($|\sigma_0|\ll|\sigma_L|$, see Ref.~\cite{Rudhardt1998}).

Ion-solvent coupling manifests itself in yet another experiment described in Ref.~\cite{Nellen2011}, 
in which the surface preference of the solvent has been measured by surface plasmon resonance.
It is reported that a hydrophilic surface ($h_0<0$) becomes less hydrophilic upon adding salt,
whereas no changes have been detected for a hydrophobic surface ($h_0>0$).
According to Eq.~\Eq{effdf} the ''external'' field $U$ for a semi-infinite system ($L\to\infty$)
acts like an additional, hydrophobic surface field 
$\delta h_0=-\int_0^\infty\d zU(z)\exp(-z/\xi)=\Delta\gamma\sigma_0\kappa\xi/(2(1+\kappa\xi))>0$
if $\Delta\gamma, \sigma_0<0$.
A hydrophilic surface becomes less hy\-dro\-phi\-lic by adding salt or for $T\to T_c$,
whereas a hy\-dro\-pho\-bic surface is influenced only weakly as $|\sigma_0|$
is small.


\paragraph{Conclusion.} We have demonstrated that even though electrolytes do not alter the
universal critical behavior of polar solvents close to their critical point, the 
ion-solvent coupling is relevant further away, provided the direct electrostatic interaction is
sufficiently weak.
The crossover from an electrostatics- to a critical Casimir-dominated behavior is expected to
occur near that temperature at which the bulk correlation length becomes comparable with
the Debye screening length.
Several experiments with monovalent ions in binary liquid mixtures can be consistently 
interpreted in terms of the present model, according to which the influence of the ions on the
order parameter can be described by an effective ``external'' field proportional to a coupling 
parameter which measures the difference between the solubility contrasts of
cations and anions in a binary solvent.
The insight gained in the present study on the effects of ion-solvent coupling may 
provide an understanding of other situations in which critical and non-critical fields
are coupled.

\begin{acknowledgments}
We thank U.\ Nellen, J.\ Dietrich, and C.\ Bechinger for many stimulating discussions.
A.G.\ is supported by MIUR within ``Incentivazione alla mobilit\`{a} di studiosi stranieri e
italiani residenti all'estero.''
M.O.\ acknowledges support by DFG-SFB TR6/N01.
\end{acknowledgments}



\end{document}